\documentclass[fleqn,usenatbib]{mnras}

\usepackage{newtxtext,newtxmath}

\usepackage[T1]{fontenc}
\usepackage{CJKutf8}
\def\oiii{[{O~\sc III}]}
\def\feii{{Fe~\sc II}}

\DeclareRobustCommand{\VAN}[3]{#2}
\let\VANthebibliography\thebibliography
\def\thebibliography{\DeclareRobustCommand{\VAN}[3]{##3}\VANthebibliography}

\usepackage{graphicx}	
\usepackage{amsmath}	

\title[jets of Narrow-line Seyfert 1 galaxies]{The jet formation mechanism of Gamma-ray Narrow-line Seyfert 1 Galaxies}

\author[Yongyun Chen et al.]{Yongyun Chen\begin{CJK*}{UTF8}{gkai}(陈永云)\end{CJK*}\thanks{E-mail: ynkmcyy@yeah.net}$^{1}$,
Qiusheng Gu\begin{CJK*}{UTF8}{gkai}(顾秋生)\end{CJK*}\thanks{E-mail: qsgu@nju.edu.cn}$^{2}$,
Junhui Fan\begin{CJK*}{UTF8}{gkai}(樊军辉)\end{CJK*}$^{3}$,
Xiaoling Yu \begin{CJK*}{UTF8}{gkai}(俞效龄)\end{CJK*}$^{1}$,
\and Nan Ding\begin{CJK*}{UTF8}{gkai}(丁楠)\end{CJK*}$^{4}$,
Xiaotong Guo \begin{CJK*}{UTF8}{gkai}(郭晓通)\end{CJK*}$^{5}$,
Dingrong Xiong\begin{CJK*}{UTF8}{gkai}(熊定荣)\end{CJK*}$^{6}$
\\
$^{1}$College of Physics and Electronic Engineering, Qujing Normal
University, Qujing 655011, P.R. China\\
$^{2}$School of Astronomy and Space Science, Nanjing University, Nanjing 210093, P. R. China\\
$^{3}$Center for Astrophysics,Guang zhou University,Guang zhou510006, China\\
$^{4}$School of Physical Science and Technology, Kunming University 650214, P. R. China\\
$^{5}$School of mathematics and physics, Anqing Normal University 246011, P. R. China\\
$^{6}$Yunnan Observatories, Chinese Academy of Sciences, Kunming 650011, China\\
}

\date{Accepted XXX. Received YYY; in original form ZZZ}

\pubyear{2022}

\begin{document}
\label{firstpage}
\pagerange{\pageref{firstpage}--\pageref{lastpage}}
\maketitle

\begin{abstract}
Under the coronal magnetic field, we estimate the maximal jet power of the Blandford–Znajek (BZ) mechanism, Blandford–Payne (BP) mechanism, and hybrid model. The jet power of the BZ and Hybrid model mechanisms depends on the spin of a black hole, while the jet power of the BP mechanism does not depend on the spin of a black hole. At high black hole spin, the jet power of the hybrid model is greater than that of the BZ and BP mechanisms. We find that the jet power of almost all gamma-ray narrow-line Seyfert 1 galaxies
($\gamma$NLS1s) can be explained by the hybrid model. However, one source with jet power $\sim0.1-1$ Eddington luminosity can not be explained by the hybrid model. We suggest that the magnetic field dragged inward by the accretion disk with magnetization-driven outflows may accelerate the jets in this $\gamma$NLS1.
\end{abstract}




\section{Introduction}
The narrow-line Seyfert 1 galaxy is a special subclass of active galactic nuclei (AGNs) defined by \cite{Osterbrock1985}. They show some extreme characteristics, such as the full width at half maximum of their Balmer line ($FWHM (H\beta) < 2000 km~s^{-1}$), \oiii/H$\beta <3$, strong \feii  ~multiplets emission, soft X-ray spectra and  rapid X-ray variability  \citep[e.g.][]{Boroson1992,Boller1996,Wang1996,Rani2017}. These narrow-line Seyfert 1 galaxies are generally considered to be radio quiet \citep[e.g.][]{Ulvestad1995, Boroson2002}. Therefore, it is surprising that the radio-loud narrow-line Seyfert 1 galaxies (RLNLS1s) was discovered for the first time \citep[e.g.][]{Remillard1986, Grupe2000, Oshlack2001, Zhou2003, Yao2021}. In particular, since the launch of Fermi Large Area
Telescope (LAT), some RLNLS1 have been found to have gamma-ray radiation \citep{Abdo2009}, which suggests that these RLNLS1s have strong relativistic jets. These RLNLS1s have a compact core-jet structure, flat radio spectra, and high brightness temperature \citep{Komossa2006, Doi2006}. It is proposed that RLNLS1s exhibit the characteristics of blazars. \cite{Yuan2008} studied a sample of 23 RLNLS1s with radio loudness larger than 100 and found that these sources display a blazar-like nature. \cite{Foschini2011} used 7 $\gamma$-ray narrow-line Seyfert 1 galaxies ($\gamma$NLS1s) to study the relation between $\gamma$NLS1s and blazars. These $\gamma$NLS1s and flat-spectrum radio quasars (FSRQs) are in the region dominated by radiation pressure. \cite{Foschini2015} found that the central engine of flat radio-loud Narrow-line Seyfert 1 galaxies (F-RLNLS1s) is similar to blazars. \cite{Sun2015} suggested that the radiation physics and the jet properties of the GeV NLS1 galaxies are similar to that of FSRQs. \cite{Paliya2019} compared the broadband parameters derived from the spectral energy distribution (SED) between 16 $\gamma$NLS1s and Fermi blazars and found that the physical properties of $\gamma$NLS1s are similar to Fermi blazars. \cite{Chen2019} studied the relation between jet and accretion for fermi blazars and F-RLNLS1s, respectively. They found that the slope of the relation between jet and accretion in Fermi blazars is similar to that of F-RLNLS1s. This result may imply that the jet formation mechanism of Fermi blazars is similar to that of RLNLS1s. \cite{Chen2021a} studied the relation between the fermi blazars sequence and $\gamma$NLS1s and found that the $\gamma$NLS1s belong to the fermi blazars sequence.        

The formation of jets has always been a hot issue in astrophysics research.
At present, there are three main theories of jet formation. First, the Blandford-Znajek mechanism (BZ): the jet extracts the rotation energy of the black hole and the accretion disk \citep{Blandford1977}. The jet power depends on the black hole mass and the spin of a black hole. \cite{Chen2021b} used a large sample of Fermi blazars to study the relation between jet power and the spin of black hole and found that the jet power depends on the spin of a black hole. Second, the Blandford-Payne mechanism (BP): the jet extracts the rotational energy of the
accretion disk \citep{Blandford1982}. The jet power depends on the luminosity of the accretion disk. This correlation has been confirmed by many authors \citep[e.g.][]{Rawlings1991, Celotti2008, Sbarrato2014, Ghisellini2010, Ghisellini2014, Zamaninasab2014, Chen2015, Mukherjee2019}. Third, the hybrid model: a mixture of the BP and BZ mechanism \citep{Meier2001,Garofalo2010}. \cite{Zhou2009} found that there is a strong relation between Lorentz factors and black hole mass, which supports that the BZ mechanism may dominate over the BP mechanism for blazars. The GeV-NLS1s have high accretion rates, which may imply that the jet of GeV-NLS1s is produced by the BP mechanism and/or BZ mechanism \citep{Zhang2015}. Some AGNs with ultra-fast outflows may imply a hybrid mechanism in these AGNs, for example, radio galaxies. \cite{Komossa2018} found that RLNLS1s have Extreme gaseous outflows, which implies that RLNLS1s may be a hybrid mechanism.            

It is generally believed that a large-scale magnetic field plays a vital role in the acceleration and collimation of jet and/or outflows \citep[e.g.][]{Pudritz2007}. The structure of the magnetic field is a key ingredient in BZ, BP, and hybrid jet models. The kinetic energy of the rotating black hole and the accretion disk is connected to the jets with the corotating large-scale magnetic field. The origin of the large-scale magnetic field passing through the accretion disk has not been well understood. Some studies show that the large-scale magnetic field of accelerating jet or outflow may be formed by the advection of a weak external field \citep{Bisnovatyi1974, Bisnovatyi1976, van1989, Spruit2005}. However, \cite{Lubow1994} found that the advection of the external field is rather inefficient in the geometrically thin accretion disk ($H/R\ll1$) due to its small radial velocity. This may imply that the field in the inner region of the disk is not much stronger than the external weak field \citep{Lubow1994}, which can not accelerate strong jets in radio-loud quasars. A few mechanisms were suggested to alleviate the difficulty of
field advection in the thin disk \cite[e.g.,][]{Spruit2005, Lovelace2009, Guilet2012, Guilet2013, Cao2013}. It has been suggested that the external field can be effectively dragged inward through the hot corona above the accretion disk, i.e., the ‘coronal mechanism’ \citep{Beckwith2009}. The radial velocity of the gas above the disk can be greater than the radial velocity of the middle plane of the disk, which partially solves the problem of inefficient field advection in the thin disk \citep{Lovelace2009, Guilet2012, Guilet2013}. Alternatively, \cite{Cao2013} suggested that if the magnetization-driven outflows remove the most angular momentum of the gas in the thin disk, the radial velocity of the disk increases significantly and, therefore, the external field in the inner region of the thin disk with the magnetization-driven outflows can be significantly enhanced.

Although there are many studies on RLNLS1s. However, there is a lack of specific research on the jet formation mechanism of $\gamma$NLS1s. Some authors have suggested that $\gamma$NLS1s (1H 0323+342) have shown the disk-corona-jet connection \citep[e.g.,][]{Paliya2014, Yao2015, Landt2017, Ghosh2018}. Thus, $\gamma$NLS1s is an ideal object for exploring the jet–corona–disk connection \citep[e.g.][]{Wilkins2015}. In this work, we consider the strength of the magnetic field of the corona and estimate the maximal jet power of BZ, BP, and hybrid jet models. We use a sample of $\gamma$NLS1s to study their jet formation mechanisms. 

\section{The Sample}
We are trying to collect a large sample of $\gamma$NLS1s with reliable redshift, black hole mass, disk luminosity, and X-ray luminosity. We consider the sample of \cite{Paliya2019}. \cite{Paliya2019} collected a sample of 16 $\gamma$NLS1s and studied their physical properties. The black hole mass and disk luminosity of 16 $\gamma$NLS1s are from the work of \cite{Paliya2019}. The accretion disk luminosity and black hole
mass are obtained by using the SED modeling \citep[see][]{Paliya2019}.  Furthermore, whenever the big blue
bump is visible at optical–UV energies, it can be attributed to
the accretion disk radiation \citep[e.g.,][]{Ghisellini2009}. By reproducing this emission with a standard disk model \citep[e.g.,][]{Shakura1973}, both black hole mass and accretion disk luminosity can be calculated. The X-ray luminosity in 0.3-10 keV comes from the work of \cite{Paliya2019}.    

\cite{Komossa2018} used the following formula to estimate the jet power of RLNLS1s \citep{Birzan2008}, 
\begin{eqnarray}
\log P_{\rm cav}=0.35(\pm0.07)\log P_{1.4} + 1.85(\pm 0.10)
\end{eqnarray}
where $P_{\rm cav}$ is in units $10^{42} erg~s^{-1}$, and $P_{1.4}$ in units $10^{40} erg~s^{-1}$. The $P_{1.4}$ is radio luminosity at 1.4 GHz. The radio luminosity is estimated by using the relation $L_{\rm \nu}=4\pi d_{\rm L}^{2}S_{{\rm \nu}}$, $S_{(\rm \nu)}= S_{\rm \nu}^{obs}(1+z)^{\alpha-1}$, where $\alpha$ is spectral index, $\alpha=0$ is assumed \citep{Donato2001, Abdo2010, Komossa2018}. Following \cite{Komossa2018}, we also use equation (1) to estimate the jet power. The data is listed in Table 1.    

\section{Magnetic field of hot Corona}
The detailed properties of the corona above the accretion disk have been extensively studied by many authors \citep[e.g.][]{Shakura1973, Galeev1979, Haardt1993, Svensson1994, Cao2009, Cao2018}. \cite{Cao2018} used relative thickness $\tilde{H}_{\rm c}=H_{\rm c}/R$ and the optical depth $\tau_{\rm c}$ to describe the corona above the accretion disk. The optical depth of the corona in the vertical direction is 

\begin{equation}
\tau_{\rm c}=\rho_{c}H_{\rm c}\kappa_{\rm T},	
\end{equation}
where $\rho_{\rm c}$ is the density of the corona, $H_{\rm c}$ is the corona thickness, and $\kappa_{\rm T}=0.4g^{-1}cm^{2}$ is the Compton scattering opacity.

The gas pressure of the corona is

\begin{equation}
p_{\rm c}=\frac{\rho_{\rm c}}{2}\left(\frac{H_{\rm c}}{R}\right)^{2}\frac{L_{*}^{2}}{R^{2}},
\end{equation}
and the $L_{*}$ is as follows,

\begin{equation}
L_{*}^{2}=L^{2}-j^{2}(E^{2}-1)
\end{equation}
where $j$ is the spin of a black hole, L is the conserved angular momentum of the gas $L=u_{\rm \phi}$, $E$ is the conserved energy $E=-u_{\rm t}$. In order to estimate the maximum jet power, $L_{*}=L_{\rm K}(r_{\rm ms})$ is used in calculating the strength of the magnetic field \citep{Cao2018}. 

\cite{Cao2018} used a parameter $\beta$ to describe the magnetic field strength in the corona, 

\begin{equation}
p_{\rm m}=\frac{B_{\rm z}^{2}}{8\pi}=\beta p_{\rm c},
\end{equation}
where $B_{\rm z}$ is the strength of the vertical component of the field, and $\beta<1$ is required for the field advected by the corona. The magnetic field strength of corona is as follows, 

\begin{equation}  
B=4.37\times10^{8}\beta^{1/2}\tau_{\rm c}^{1/2}\tilde{H}_{\rm c}^{1/2}m^{-1/2}r^{-3/2}L_{*}^{2}~Gauss
\end{equation}

\begin{eqnarray}
r = \frac{Rc^{2}}{GM_{\rm bh}},  m = \frac{M_{\rm bh}}{M_{\odot}}, \tilde{H}_{\rm c} = \frac{H_{\rm c}}{R}	
\end{eqnarray}
where $H_{\rm c}$ is the corona thickness (see \cite{Cao2018}, for the details).   

\begin{table*}
	\caption{The sample of $\gamma$NLS1s}
	\centering
	\label{table1}
	\setlength{\tabcolsep}{3.5mm}{
		\begin{tabular}{llllllllccccccrrrrr} 
			\hline
			\hline
			Name & Redshift & log$(M/M_{\odot})$  & log$L_{disk}$ & $f_{\nu}$  & log$P_{jet}$ & log$L_{X}$ & log$\lambda$  &  Disk \\
			(1)    &  (2)  & (3)   & (4)  &  (5) &  (6) &  (7) & (8) & (9) \\
			\hline
1H 0323+342	&	0.061	&	7.3	&	45.3	&	613.5	&	44.163	&	44.432 	&	-1.114 	&	SS	\\
SBS 0846+513 	&	0.584	&	7.59	&	44.43	&	266.3	&	44.799	&	45.002 	&	-2.274 	&	ADAF+SS	\\
CGRaBS J0932+5306 	&	0.597	&	8	&	45.7	&	481.6	&	44.896	&	44.842 	&	-1.414 	&	SS	\\
GB6 J0937+5008 	&	0.275	&	7.56	&	43.71	&	166.6	&	44.461	&	44.892 	&	-2.964 	&	ADAF	\\
PMN J0948+0022 	&	0.585	&	8.18	&	45.7	&	69.5	&	44.594	&	45.792 	&	-1.594 	&	SS	\\
TXS 0955+326 	&	0.531	&	8.7	&	46.23	&	1247.1	&	44.998	&	44.852 	&	-1.584 	&	SS	\\
FBQS J1102+2239 	&	0.453	&	7.78	&	44	&	2.5	&	43.998	&	44.542 	&	-2.894 	&	ADAF	\\
CGRaBS J1222+0413 	&	0.966	&	8.85	&	46.18	&	800.3	&	45.155	&	46.132 	&	-1.784 	&	SS	\\
SDSS J124634.65+023809.0 	&	0.362	&	8.48	&	44.48	&	35.7	&	44.325	&	43.902 	&	-3.114 	&	ADAF	\\
TXS 1419+391	&	0.49	&	8.48	&	45.34	&	85.7	&	44.563	&	44.762 	&	-2.254 	&	ADAF+SS	\\
PKS 1502+036	&	0.407	&	7.6	&	44.78	&	394.8	&	44.733	&	44.402 	&	-1.934 	&	SS	\\
TXS 1518+423	&	0.484	&	7.85	&	44.7	&	138.4	&	44.629	&	44.572 	&	-2.264 	&	ADAF+SS	\\
RGB J1644+263	&	0.145	&	7.7	&	44.48	&	128.4	&	44.205	&	44.072 	&	-2.334 	&	ADAF+SS	\\
PKS 2004-447	&	0.24	&	6.7	&	43	&	471	&	44.57	&	44.062 	&	-2.814 	&	ADAF	\\
TXS 2116-077	&	0.26	&	7.2	&	44	&	96.1	&	44.357	&	43.892 	&	-2.314 	&	ADAF+SS	\\
PMN J2118+0013 	&	0.463	&	7.77	&	44.94	&	147.9	&	44.625	&	44.522 	&	-1.944 	&	SS	\\
			\hline
	\end{tabular}}
	\footnotesize{Columns (1) is the name of sources; Columns (2) is redshift; Columns (3) is the black hole mass; Columns (4) is the disk luminosity; Columns (5) is the 1.4 GHz radio flux in units mjy; Columns (6) is the jet power in units $erg~s^{-1}$; Columns (7) is the X-ray luminosity in 2-10 keV; Columns (8) is the dimensionless accretion rates; Columns (9) is the disk.}
\end{table*}
 
\section{Jet formation model}
\subsection{The BZ jet model}
The jet power extracted from a rapidly rotating black hole is \citep[e.g.][]{MacDonald1982, Ghosh1997} 

\begin{equation}
P_{\rm jet}^{\rm BZ} = \frac{1}{32}\omega_{\rm F}^{2}B_{\perp}^{2}R_{\rm H}^{2}j^{2}c	
\end{equation} 
where $B_{\perp}$ is the magnetic field strength at the black hole horizon,  $B_{\perp}\simeq B$. $R_{\rm H}=[1+(1-j^{2})^{1/2}]GM_{\rm BH}/c^{2}$ is the horizon radius. $\omega_{\rm F}\equiv\Omega_{\rm F}(\Omega_{\rm H}-\Omega_{\rm F})/\Omega_{\rm H}^{2}$ is the angular velocity of the field lines $\Omega_{\rm F}$ relative to that of the hole. $\omega_{\rm F}=1/2$ is used when estimating the maximum BZ jet power \citep[e.g.][]{MacDonald1982, Ghosh1997}. Substituting equation (6) into equation (8), we can get the the maximum BZ jet power. 

\subsection{The BP jet model}
The jet power generated by the BP mechanism can be calculated using the following formula \citep{Livio1999, Cao2018}, 

\begin{equation}
P_{\rm jet}^{\rm BP}\sim \frac{BB_{\rm \phi}^{s}}{2\pi}R_{\rm j}\Omega\pi R_{\rm j}^{2}
\end{equation}
where $R_{\rm j}$ is the radius of the jet formation region in the corona, $\Omega$ is the angular velocity of the gas in the corona. The $B_{\rm \phi}^{\rm s}$ is the azimuthal component of magnetic field at the corona surface, $B_{\rm \phi}^{\rm s}=\xi_{\phi}B$. The ratio $\xi_{\phi} \leq 1$  is required \citep{Livio1999}. Substituting equation (6) into equation (9), the jet power of BP can be derived as follows

\begin{equation}
P_{\rm jet}^{BP}\simeq 3.13\times10^{37}\xi_{\phi}\tilde{\Omega}r_{\rm j}^{-1/2}m\beta\tau_{\rm c}\tilde{H_{\rm c}}~erg~s^{-1}.
\end{equation}
Most of the gravitational force can be released in the inner region of the accretion disk within the radius $\sim2R_{\rm ms}$ \citep{Shakura1973}. $R_{\rm j}=2R_{\rm ms}$ is adopted in the estimates of BP jet power. The $R_{ms}$ is defined by the following formula,

\begin{eqnarray}
	R_{\rm ms} = R_{\rm G}\{3+Z_{2}-\left[(3-Z_{1})(3+Z_{1}+2Z_{2})\right]^{1/2}\},\nonumber \\
	Z_{1}\equiv1+(1-a^{2})^{1/3}\left[(1+a)^{1/3}+(1-a)^{1/3}\right], \nonumber \\
	Z_{2}\equiv(3a^{2}+Z_{1}^{2})^{1/2},\nonumber \\
	R_{\rm G} =\frac{GM_{\rm bh}}{c^{2}}.
\end{eqnarray}

\begin{figure}
	\includegraphics[width=9.0cm,height=8.5cm]{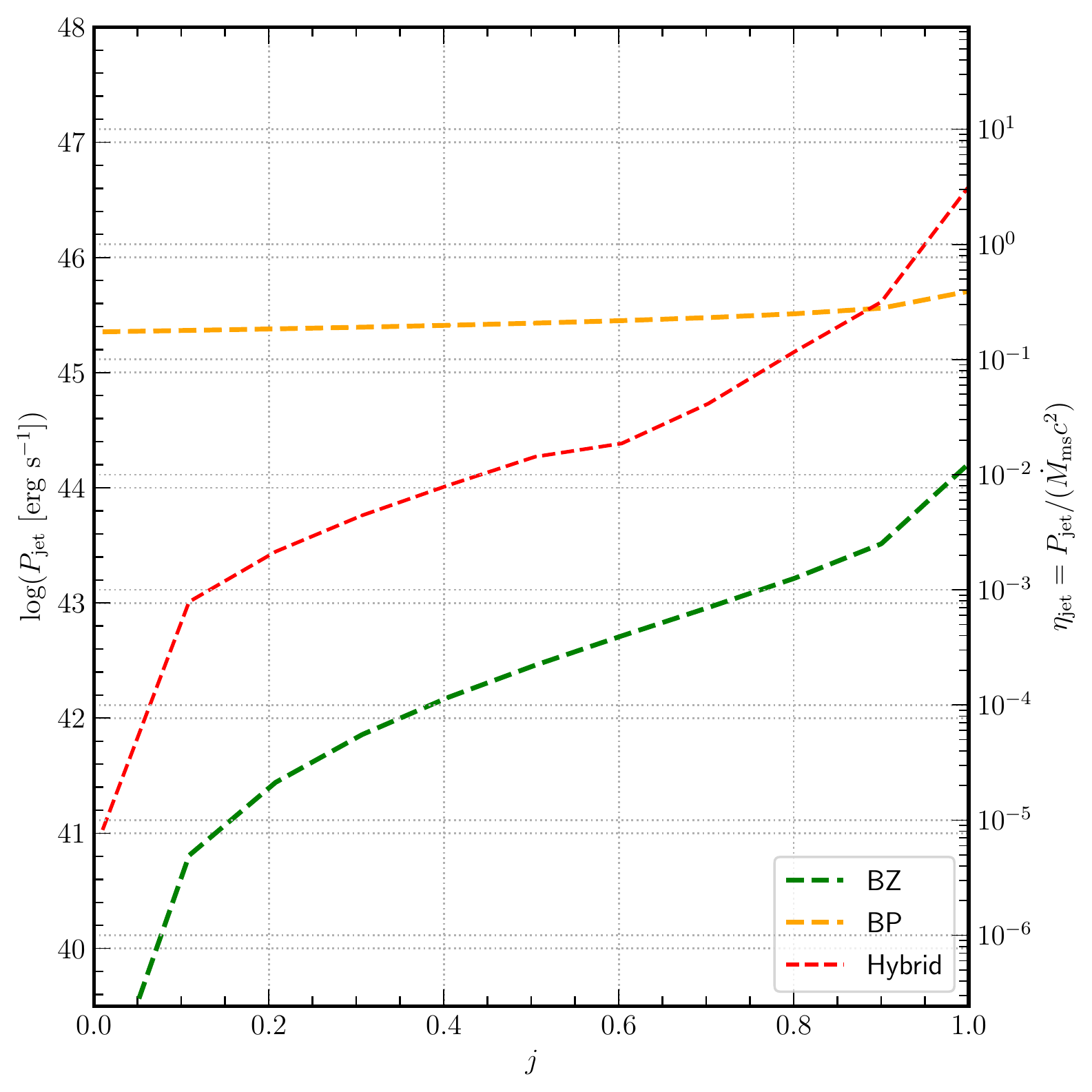}
	\caption{The black hole spin versus jet power. The orange dashed line is jet power $P_{\rm jet}^{\rm BP}$ extracted from a standard accretion disk. The green dashed line is jet power $P_{\rm jet}^{\rm BZ}$ extracted from a rapidly spinning black hole. The red dashed line is Hybrid jet model.}
	\label{figure1}
\end{figure}

\subsection{Hybrid model}
The hybrid model is that jets are produced by the combined
effort of the BZ and BP mechanism. In the case of a thin accretion disk, the total jet power for the hybrid model is given by \citep{Garofalo2009, Garofalo2010} 

\begin{equation}
P_{\rm jet}^{\rm Hybrid} = 2\times10^{47} erg~s^{-1}\alpha f^{2}\left(\frac{B_{\rm pd}}{10^{5}G}\right)^{2}m_{9}^{2}j^{2},
\end{equation} 
$B_{\rm pd}\simeq B$, $m_{9}$ is the black hole mass in units of $10^{9}M_{\odot}$, the functions $\alpha$ and $f$ capture the effects on BP and BZ powers, respectively, from the numerical solution in the context
of the Reynolds conjecture \citep{Garofalo2009}  
\begin{eqnarray}
\alpha=\delta\left(\frac{3}{2}-j\right)	
\end{eqnarray}
and 
\begin{equation}
\begin{split}
f=-\frac{3}{2}j^{3}+12j^{2}-10j+7-\frac{0.002}{(j-0.65)^{2}}+\frac{0.1}{(j+0.95)}\\+\frac{0.002}{(j-0.055)^{2}}.	
\end{split}
\end{equation}
where a conservative value for $\delta$ of about 2.5 is adopted but our ignorance of how jets couple the BZ and BP components restrict our ability to specify it and suggests that it might well be larger by an order of magnitude or more \citep{Garofalo2010}. While $\alpha$ can be thought of as the parameter that determines the effectiveness of the BP jet as a function of spin, $f$ captures the enhancement of the disk thread field to the black hole, both of which are within the Reynolds conjecture. The parameter $\delta$
determines the effective contribution of BP to overall jet power,
with larger values shifting jet power efficiency more towards the
retrograde regime.

\section{RESULTS AND DISCUSSION}
The magnetic pressure is required to be lower than the gas pressure in the corona when the magnetic field is dragged inward by the corona, i.e. $\beta<1$. In order to calculate the maximum jet power, we use $\omega_{\rm F}=1/2$ \citep{Ghosh1997}, $\beta=1$, $\xi_{\phi}=1$ and $\tilde{\Omega}=1$, while the typical values of the corona parameters, $\tau_{\rm c}=0.5$ and $\tilde{H_{\rm c}}=0.5$ are used in the estimates \citep{Cao2009, Cao2018}. The relationship between the maximum jet power and the spin parameters of the black hole is shown in Figure 1. 

In Figure 1, the jet power of BZ and hybrid models changes similarly with the spin of the black hole. The slight difference is that the jet power of the hybrid model is higher than that of
the BZ model. The sharp drop in the BZ case reflects the drop
in the rotational energy of the black hole, which can be used
to enhance the jet. In the case of hybrid, the decline is limited
by the contribution from the accretion disk. The most important
point is that the jet power of the BZ and hybrid model strongly
depends on the spin of a black hole. However, the jet power of
BP does not depend on the spin of the black hole. The jet power
of BP is always higher than that of the BZ and hybrid model when
the spin of a black hole is less than 0.9. However, the jet power
of the hybrid model is greater than that of BP when the spin of
the black hole is greater than 0.9.

The right-hand axis of Figure 1 shows the jet efficiency factor, defined as $\eta_{\rm jet}\equiv P_{\rm jet}/\dot{M}_{\rm ms}c^{2}$. Because the jet power is related to $\dot{M}_{\rm ms}$, the corresponding jet efficiency only depends on the spin of the black hole ($j$) and viscosity parameters. From figure 1, we find that the hybrid models have high jet efficiency when the spin of the black hole is greater than 0.9. The result suggests that the magnetic field of hot corona may explain observations of AGNs with high jet efficiency $\eta \approx few \times 100~per~cent$.  

\begin{figure}
	\includegraphics[width=8.5cm,height=8.5cm]{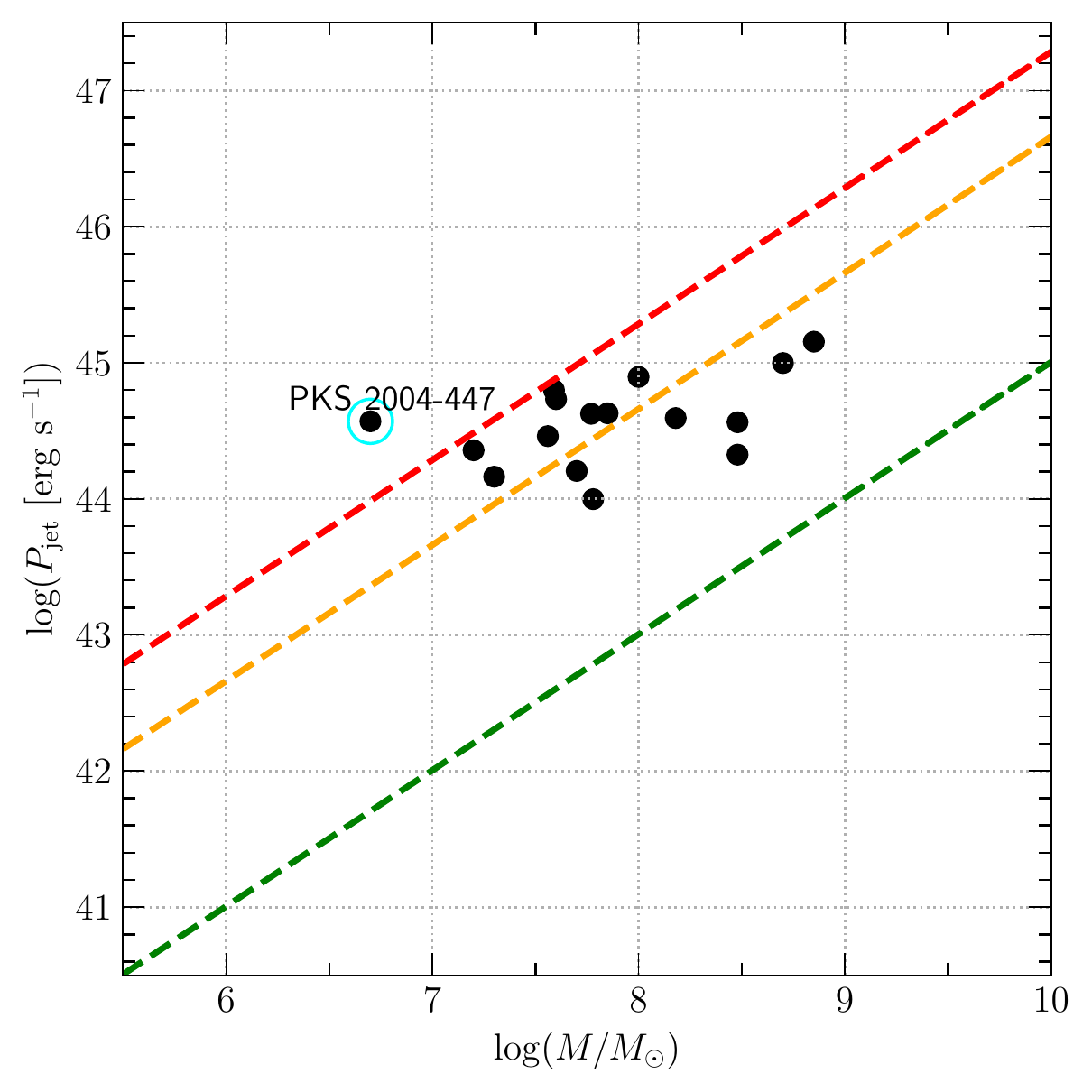}
	\caption{The black hole mass versus maximal jet power. The orange dashed line is maximal jet power $P_{\rm jet}^{\rm BP}$ extracted from a standard accretion disk. The green dashed line is maximal jet power $P_{\rm jet}^{\rm BZ}$ extracted from a rapidly spinning black hole. The red dashed line is Hybrid jet model. The black circle is $\gamma$NLS1s. The spin of black hole $j=0.95$ are adopted. The cyan circle marks PKS 2004-447.}
	\label{figure2}
\end{figure}

The relation between jet power and black hole mass is shown
in Figure 2. The red dashed line is a hybrid jet model. The orange dashed line is the BP jet model. The green dashed line is
the BZ jet model. The black circle dot is $\gamma$NLS1s. 
From figure 2, we find that the hybrid jet model has high jet power than that of both BP and BZ jet power. About 55\% sources are above both the orange and green lines, which implies that BP and BZ mechanisms cannot explain the jet power of these sources. However, we can see that almost all sources are located below the red line, which shows that the jet power of these sources can be explained by the hybrid jet model. Here, the jet power of only one source (PKS 2004-447) is above the red line. There are two possible explanations for one source. One explanation is that the one source has a strong beaming effect, and it has high jet power (PKS 2004-447). \cite{Schulz2016} found a compact core by using the Very-Long-Baseline Interferometry (VLBI) observation in PKS 2004-447. The PKS 2004-447 is a core-dominated, one-sided parsec-scale jet with a strong beaming effect. \cite{Orienti2015} also found that the PKS 2004-447 have a high Doppler factor ($D_{factor}=30$). These results suggest that the PKS 2004-447 have a strong beaming effect. Thus, it has high jet power. Another explanation is that the one source may require other jet models, such as the accretion disk with magnetization-driven outflows models.   

\begin{figure}
	\includegraphics[width=8.5cm,height=8.5cm]{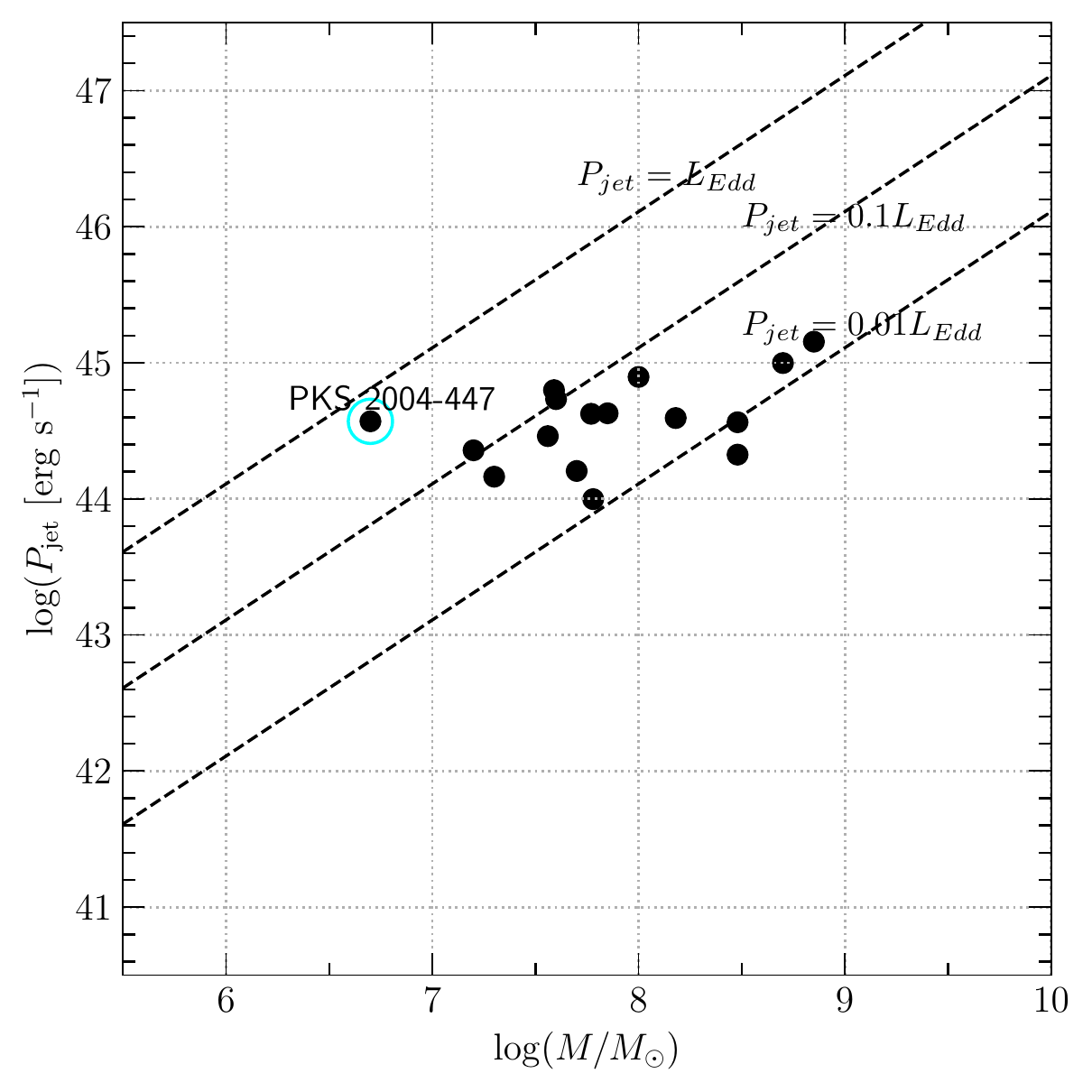}
	\caption{The black hole mass versus jet power. The dashed lines are $P_{\rm jet}/L_{\rm Edd}=0.01, 0.1, 1$, respectively. The cyan circle marks PKS 2004-447.}
	\label{figure3}
\end{figure}

In Figure 3, we present the results as in Figure 2, but the dash lines are 
$P_{\rm jet}/L_{\rm Edd}=0.01, 0.1, 1$, respectively. All sources have jet power less than 1$L_{\rm Edd}$. There are about two sources located at $\sim0.1-1L_{\rm Edd}$. Some authors suggested that the magnetic field dragged inwards by the accretion disk with magnetization-driven outflows may accelerate the jets in AGNs \citep[e.g.][]{Cao2014, Li2014, Cao2016, Li2019b, Cao2021}. In this scenario, the hot corona/gas above the accretion disk may help to enhance the jet/outflow \citep{Li2021}. Most of the angular momentum of the accretion disk is removed by magnetization-driven outflows, and the magnetic field will be enhanced than the thin disk without outflow \citep{Cao2013}.  \cite{Cao2018} suggested that the magnetic field dragged inward by the accretion disk with magnetization-driven outflows may help explain the sources with jet power $\sim0.1-1L_{\rm Edd}$ (or even higher). The outflow can be enhanced by the radiation pressure of the accretion disk \citep[e.g.][]{Bisnovatyi-Kogan1977, Shlosman1985, Murray1995}. This scenario is very effective for the jet of a source with a high mass accretion rate, for example, RLNLS1s and radio galaxies. Radiation pressure may play an important role in magnetization-driven outflows \citep{Cao2014}. Some authors have found that the accretion disk of $\gamma$NLS1s is dominated by radiation-pressure \citep[e.g.,][]{Foschini2011, Chen2019}. Our results may imply that the jet power of the one source may be explained by the magnetization-driven outflows model.     

We also note that the NLS1 has a high accretion rate, which may imply different accretion disk models. \cite{Wang2002} defined the line  dimensionless accretion rate form as 

\begin{eqnarray}
\lambda = \frac{L_{\rm lines}}{L_{\rm Edd}}, L_{\rm lines}=\xi L_{\rm disk}.
\end{eqnarray}
where $\xi\approx0.1$ \citep{Netzer1990}. \cite{Wang2002} got the relation between $\lambda$ and the dimensionless accretion rate ($\dot{m}$) for an optically thin ADAF, 

\begin{equation}
\dot{m}=2.17\times10^{-2}\alpha_{0.3}\xi_{-1}^{-1/2}\lambda_{-4}^{1/2}.
\end{equation}
where $\alpha_{0.3}=\alpha/0.3$, $\xi_{-1}=\xi/0.1$, and $\lambda_{-4}=\lambda/10^{-4}$. A necessary condition for the
presence of an optically thin ADAF is $\dot{m}\leq\alpha^{2}$ \citep{Narayan1998}. Equation (16) can then be rewritten as

\begin{equation}
\lambda_{1}=1.72\times10^{-3}\xi_{-1}\alpha_{0.3}^{2}.
\end{equation}
Optically thin ADAFs require $\lambda<\lambda_{1}$. The optically thick, geometrically thin disk obey 

\begin{equation}
\dot{m} = \frac{L_{\rm line}}{\xi L_{\rm Edd}} = 10\xi_{-1}^{-1}\lambda.
\end{equation} 
The accretion disk is standard thin disk (SS) when accretion rate satisfy $1>\dot{m}\geq\alpha^{2}$ \citep[e.g][]{Shakura1973}. Equation (18) can then be rewritten as

\begin{equation}
	\lambda_{2}=9.0\times10^{-3}\xi_{-1}\alpha_{0.3}^{2}.
\end{equation}
A standard thin disk can exist when gives $\lambda\geq\lambda_{2}$. It is a super Eddington accretion (SEA) when the accretion rate reaches $\dot{m}\geq1$, we have 

\begin{equation}
\lambda_{3}=0.1\xi_{-1}.
\end{equation}   
A slim disk requires $\lambda\geq\lambda_{3}$ \citep{Wang2002, Wang2003}. 
Interestingly, in the transition region between $\lambda_{1}$ and $\lambda_{2}$, the accretion flow may be in a mixed state where the standard disk and ADAF coexist. Several authors have discussed the possibility of mixed states in AGN accretion disks \citep[e.g.,][]{Quataert1999, Rozanska2000, Ho2000}. A transition from an SS disk to an ADAF is possible
\citep[e.g.,][]{Gu2000}, perhaps via evaporation \citep{Liu1999}. The
transition radius depends on the accretion rate, the black hole mass,
and viscosity. The disk structure, however, is complicated in
such a regime.

\begin{figure}
	\includegraphics[width=8.5cm,height=8.5cm]{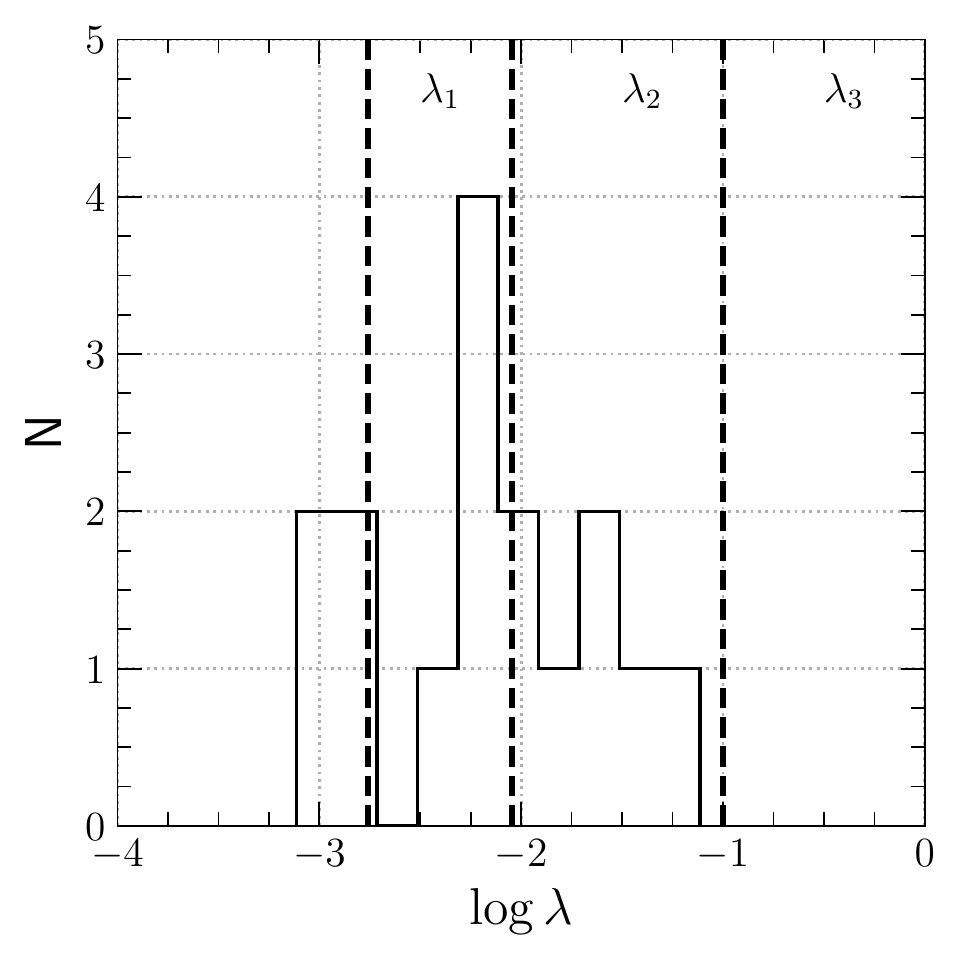}
	\caption{The distribution of $\lambda$. The distribution of $\lambda$ is divided into four regions, corresponding to different states of the accretion disks:(1) $\lambda<\lambda_{1}$ (pure ADAF); (2) $\lambda_{1}\leq\lambda\leq\lambda_{2}$ (ADAF+SS); (3) $\lambda_{2}\leq\lambda\leq\lambda_{3}$ (SS); (4) $\lambda\geq\lambda_{3}$ (SEA).}
	\label{figure4}
\end{figure}

The distribution of the three critical values of $\lambda$ define four regimes in accretion states are shown in Figure 4. The average values of $\lambda$ for $\gamma NLS1s$ are $\langle \log\lambda \rangle = -2.16\pm0.56$. From figure 4, we find that the distribution of $\log\lambda$ is from -3.2 to -1.1. These $\gamma$NLS1s do not have super Eddington accretion and do not need slim disk interpretation. From Figure 2, almost all $\gamma$NLS1s can be explained by the standard disk. We also note that values of $\lambda_{1}$ and $\lambda_{2}$ are sensitive to $\alpha$. The most likely value of $\alpha$ is 0.3 \citep{Narayan1998}. However, if $\alpha=0.1$, it will follow that $\lambda_{1}=1.91\times10^{-4}\xi_{-1}\alpha_{0.1}^{2}$ and $\lambda_{2}=1.0\times10^{-3}\xi_{-1}\alpha_{0.1}^{2}$. All the disks listed as hybrid models of SS and ADAF will be SS disks in Table 1. In any case, this
does not affect the conclusion that most of the objects have the
standard accretion disks.   

It is well known that the luminosities of typical NLSys are around the Eddington value. However, we find that this sample has a lower Eddington ratio. The possible reason is due to the black hole mass. Previously, the black hole mass derived from the optical spectroscopy is argued to be underestimated, due to the radiation pressure effect in highly accreting systems and also possibly due to a flat BLR geometry \citep{Decarli2008, Marconi2008}. The latter effect is proposed to be more pronounced in the sources with a pole-on view, typically blazars and $\gamma$NLS1s. On the other hand, the disk modeling approach is free from these issues and relies mainly on the visibility of the big blue bump. \cite{Calderone2013} modeled the SED of a sample of radio-loud NLSy1 galaxies with a standard \cite{Shakura1973} accretion disk spectrum and estimated a consistently higher black hole mass than that derived from the optical spectroscopy. Because the black hole mass of our sample comes from the SED model, these gamma-ray Narrow-line Seyfert 1 galaxies may have a low Eddington ratio.

\cite{Meier2001} considered the SS disk and ADAF extracting energy from the spin of the black hole and the disk itself. His model uses an intermediate region solution for SS disks and a self-similar solution for ADAF; it takes into account Schwarzschild and Kerr's black holes. In this model, if the Kerr black holes with the same mass and accretion rate are $M_{\rm BH}=10^{9}M_{\odot}$ and $\dot{M}=0.1$, the jet power of the Schwarzschild black hole is only is $P_{\rm jet}=10^{41.7}\rm erg~s^{-1}$, and a Kerr black hole with the same mass and accretion rate is $P_{\rm jet}=10^{42.7}\rm erg s^{ -1}$. The current data does not support this model, as many sources have SS disks that support higher luminosity. \cite{Wang2003} considered the jet power of slim disk with hot coronae, $P_{\rm jet}\approx8.0\times10^{37}M_{\rm BH}ln((1-f)\dot{m}/50) \rm erg~s^{-1}$ ($2.5\ll\dot{m}\leq100$). If the black hole mass, accretion rate and free parameter $f$ are $M_{BH}=10^{9}M_{\odot}$, $\dot{m}=100$, and $f=0$, the maximal jet power is $P_{\rm jet}=10^{46.74}\rm erg~s^{-1}$. All $\gamma$NLS1s can be explained by the slim disk with hot coronae.           

We know that the fraction of the corona to the disk emission increases with the increasing Eddington ratio. The properties of the coronae may be related to their hard X-ray emission. If the coronal jet model is indeed at work in $\gamma$NLS1s, one may expect a correlation between radio and hard X-ray emission. We test this correlation. In order to get consistent hard X-ray luminosity in 2-10 keV, we roughly convert X-ray luminosity in 0.3-10 keV, $L_{X,0.3-10 keV}$, to the hard X-ray luminosity with a mean energy spectral index of $\alpha_{\rm X}=1 (f_{x}\propto E^{-\alpha_{X}})$, $L_{X,2-10keV}=0.46L_{X,0.3-10keV}$. This is consistent with the general features of the theoretical calculations for the X-ray spectra from the disk corona \citep[e.g.,][]{Nakamura1993}. Figure 5 shows the relation between radio luminosity and hard X-ray luminosity. The Pearson analysis shows that there is a significant correlation between radio luminosity and hard X-ray luminosity ($r=0.61, P=0.01$, significant correlation P$<$0.05 confidence level). The Spearman correlation coefficient and significance level are $r=0.58$ and $P=0.02$. The Kendalltau correlation coefficient and significance level are $r=0.38$ and $P=0.04$. These two tests also show a significant correlation.       

\begin{figure}
	\includegraphics[width=8.5cm,height=8.5cm]{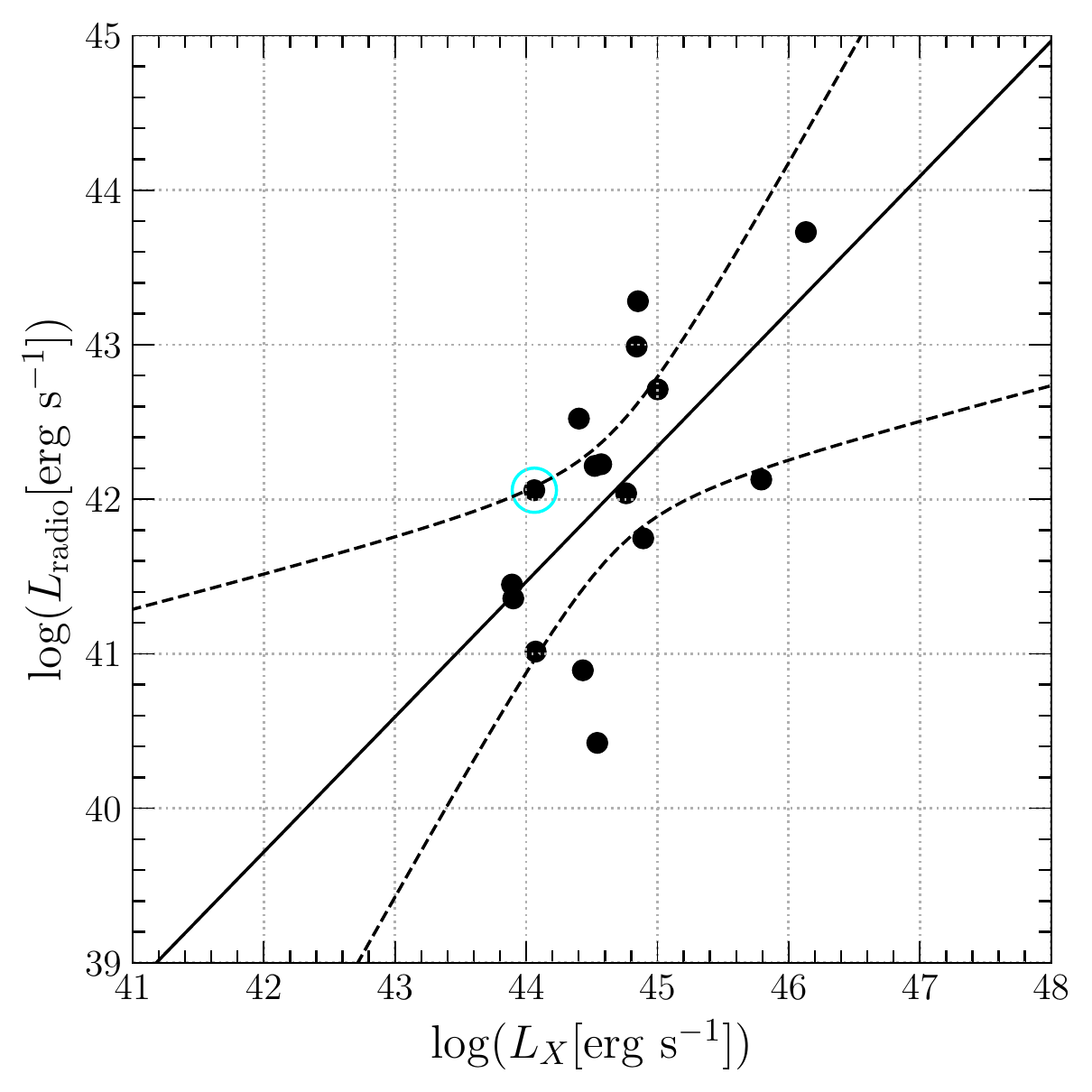}
	\caption{Relation between radio luminosity and hard X-ray luminosity for $\gamma$NLS1s. The solid lines correspond to the best-fit linear models obtained with the symmetric least-squares fit. The dashed lines indicate 2$\sigma$ confidence band. The cyan circle marks PKS 2004-447.}
	\label{figure5}
\end{figure}

\begin{figure}
	\includegraphics[width=8.5cm,height=8.5cm]{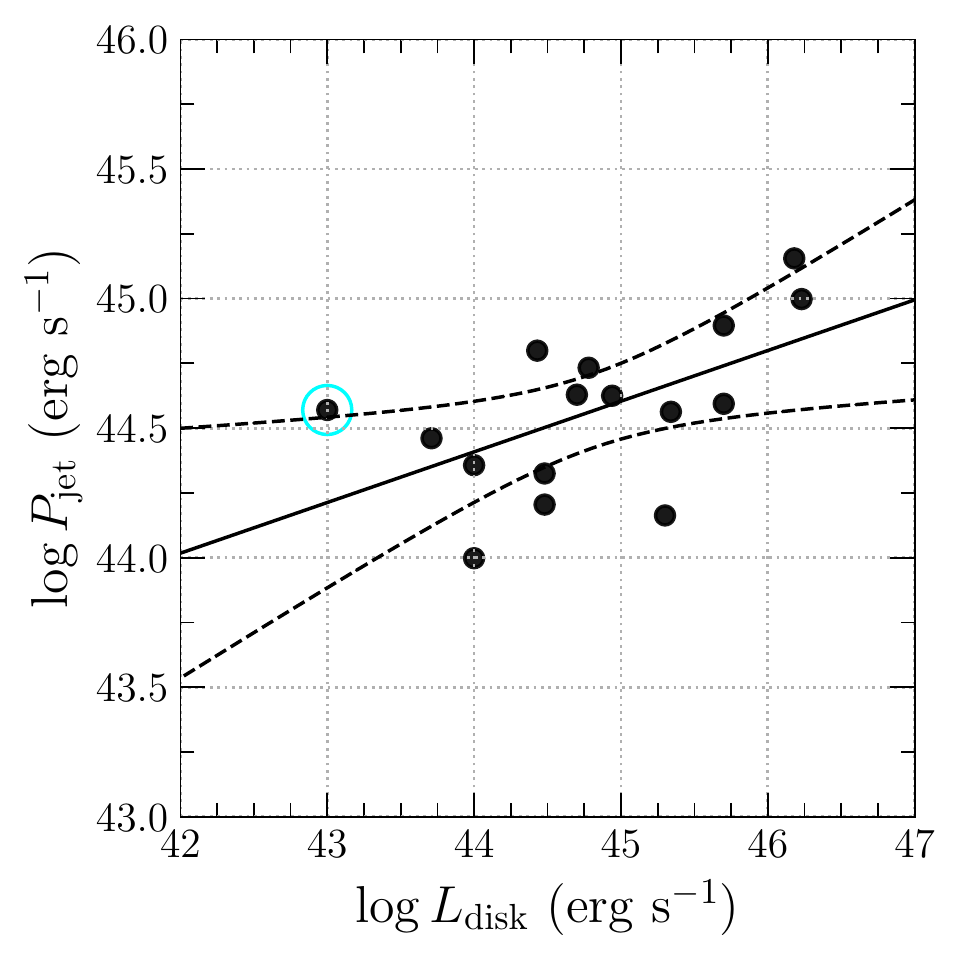}
	\caption{Relation between jet power and accretion disk luminosity for $\gamma$NLS1s. The solid lines correspond to the best-fit linear models obtained with the symmetric least-squares fit. The dashed lines indicate 2$\sigma$ confidence band. The cyan circle marks PKS 2004-447.}
	\label{figure6}
\end{figure}

It is generally believed that there is a close connection between jet power and accretion disk luminosity for radio galaxies/AGNs. We test this correlation for $\gamma$NLS1s. The relation between jet power and accretion disk luminosity for $\gamma$NLS1s is shown in Figure 6. The Pearson analysis shows that there is a significant correlation between jet power and accretion disk luminosity ($r=0.56, P=0.02$). The Spearman correlation coefficient and significance level are $r=0.55$ and $P=0.03$. The Kendalltau correlation coefficient and significance level are $r=0.36$ and $P=0.04$. These two tests also show a significant correlation. These results show that there is a close connection between jet and accretion.      

\section{SUMMARY}
We consider the magnetic field in the corona and calculate the jet power of the BZ mechanism, BP mechanism, and hybrid model. The jet power of the BZ mechanism and hybrid model depends on the spin of a black hole, while the jet power of the BP mechanism has no obvious relationship with the spin of a black hole. At high a spin of the black hole, the jet power of the hybrid model is higher than that of the BZ mechanism and BP mechanism. 

We find that the jet power of 99\% $\gamma$NLS1s is lower than that of the hybrid model. This result shows that the hybrid model can explain almost all the jet power of $\gamma$NLS1s. However, the jet power of only one source is located at $\sim0.1-1L_{\rm Edd}$. \cite{Cao2013} suggested that most of the
angular momentum of the disk can be removed by the
magnetization-driven outflows, and therefore the radial velocity of the disk increases significantly. Because the most angular momentum of the
disk is carried away by the outflows, the gas in the disk rapidly
accretes onto the black hole. Thus, the magnetic field dragged inward by the accretion disk with magnetization-driven outflows may accelerate the jet in one source. 

\section*{Acknowledgements}
We thank the referee and Editor for the helpful comments and
suggestions that improved the presentation of the manuscript
substantially. Yongyun Chen thanks for the financial support from the National Natural Science Foundation of China (No. 12203028). This work was support from the research project of Qujing Normal University (2105098001/094). This work is supported by the youth project of Yunnan Provincial Science and Technology Department (202101AU070146, 2103010006). 
This work is supported by the National Natural Science Foundation of China (NSFC 11733001). This work is supported by
the National Key Research and Development Program of
China (No. 2017YFA0402703) and the National Natural
Science Foundation of China (Grant Nos. 11733002
and 11773013).Nan Ding thanks for the financial support from the National Natural Science Foundation of China (No. 12103022) and the Special Basic Cooperative Research Programs of Yunnan Provincial Undergraduate Universities'Association (No. 202101BA070001-043).
\section*{Data Availability}
All the data used here are available upon reasonable request. All datas are in Table 1. 

\bibliographystyle{mnras}
\bibliography{example} 






\bsp	
\label{lastpage}
\end{document}